\documentclass[twocolumn]{autart}

\usepackage{graphicx}
\usepackage{amsmath}
\usepackage{amsfonts}
\usepackage{amssymb}
\usepackage{subcaption}
\usepackage{float}
\usepackage{algpseudocode}
\renewcommand{\algorithmicrequire}{\textbf{Input:}}
\renewcommand{\algorithmicensure}{\textbf{Output:}}
\usepackage{multirow}
\usepackage{enumerate}
\usepackage{tikz}
\usetikzlibrary{shapes.geometric, arrows}
\allowdisplaybreaks
\usepackage{booktabs}
\newcommand{\RN}{{\rm I\!R}}

\begin{document}
\begin{frontmatter}

\title{Hybrid zonotopes: a new set representation for reachability analysis of mixed logical dynamical systems \thanksref{footnoteinfo}} 

\thanks[footnoteinfo]{This material is based upon work supported by the National Science Foundation Graduate Research Fellowship under Grant No. DGE-1333468. This paper was not presented at any IFAC meeting. Corresponding author J. P. Koeln. Tel. +1 972-883-4649. Fax +1 972-883-4659.}

\author[Purdue]{Trevor J. Bird}\ead{bird6@purdue.edu}, 
\author[PennState]{Herschel C. Pangborn}\ead{hcpangborn@psu.edu},  
\author[Purdue]{Neera Jain}\ead{neerajain@purdue.edu}, \author[UTDallas]{Justin P. Koeln}\ead{justin.koeln@utdallas.edu}    

\address[Purdue]{School of Mechanical Engineering, Purdue University, IN, USA}
\address[PennState]{Department of Mechanical Engineering, The Pennsylvania State University, PA, USA}
\address[UTDallas]{Department of Mechanical Engineering, University of Texas at Dallas, TX, USA} 
          

\begin{abstract} 
This article presents a new set representation named the \textit{hybrid zonotope} that is equivalent to the union of $2^N$ constrained zonotopes---convex polytopes---through the addition of $N$ binary zonotope factors.
The major contribution of this manuscript is a closed-form solution for exact forward reachable sets of discrete-time, linear hybrid systems modeled as mixed logical dynamical systems. 
The proposed approach captures the worst-case exponential growth in the number of convex sets required to represent the nonconvex reachable set while exhibiting only linear growth in the complexity of the hybrid zonotope set representation.
Redundancy removal techniques are provided that leverage binary trees to store the combinations of binary factors of the hybrid zonotope that map to nonempty convex subsets. Numerical examples show the hybrid zonotope's ability to compactly represent nonconvex reachable sets with an exponential number of features.
Furthermore, the hybrid zonotope is shown to be closed under linear mappings, Minkowski sums, generalized intersections, and halfspace intersections.
\end{abstract}

\end{frontmatter}
\section{Introduction}\label{sec-introduction}

Hybrid system theory has found increased use for modeling and control synthesis due to its ability to capture the mixed continuous and discrete dynamics exhibited by many engineered systems \cite{alur_algorithmic_1995}. While providing a powerful tool, the analysis and control of hybrid systems is inherently complex. Even in the case of linear hybrid systems, basic properties such as stability and controllability may not be easily determined from the system model \cite{blondel_complexity_1999,liberzon_switching_2003}. 
Thus, hybrid systems under closed-loop control may not exhibit the intended behavior under certain operating conditions.
Set-based methods for reachability analysis and safety verification are often deployed when certain properties of a system, such as safety or performance, must be guaranteed. 
These methods are well studied for linear time invariant systems using convex sets, for which multiple representations exist \cite{blanchini_set-theoretic_2015-4}. However, the application of set-based methods to nonlinear and hybrid systems are nonconvex. The reader is directed to the review papers \cite{chen_hamiltonjacobi_2018,althoff_set_2021} and the references therein for detailed discussion on the state-of-the-art. 

In the case of linear hybrid systems, nonconvexity arises in reachable sets due to discrete inputs, switching of dynamic subsystems, and reset maps. The reachable set may be determined by partitioning the state space into a collection of closed convex sets, often referred to as locations separated by guards \cite{alur_algorithmic_1995}. Using a finite collection of convex sets and applying techniques developed for linear systems, the reachable set may be found by iteratively propagating the appropriate linear dynamics within each location. 
However, when an intersection with a guard occurs or an uncertain discrete input is applied, the reach set branches, resulting in a worst-case exponential growth in the number of convex sets required to represent the reachable space as their implicit union \cite{althoff_set_2021}.
This approach is frequently computationally intractable. 

To avoid exponential growth in set representation complexity, researchers often approximate the true reachable set, given by the implicit union of a finite number of convex sets, by a reduced number of convex sets. One such method propagates the dynamics of the system by branching along each guard, then uses clustering methods to over-approximate groups of convex sets by fewer convex sets \cite{frehse_flowpipe_2013}.
This approach provides computational efficiency at the cost of conservatism in the reachable set itself, although the specific trade-off is application-dependent.
Another approach is to search each region of the partitioned state space individually and then over-approximate transitions along the guards \cite{althoff_computing_2010}.
This approach is computationally efficient as it only propagates one convex set at a time and avoids unnecessary error by only over-approximating nonconvex sets along guard set intersections. However, it is not guaranteed to converge when the reach set intersects a guard partially without fully transitioning into another location. 

While useful and efficient, existing approaches that rely on over-approximations are only valid for safety verification and avoiding unsafe regions in robust control. Furthermore, the error associated with such over-approximations may be large and difficult to quantify, thus resulting in conservative results at best, and trivial solutions at worst \cite{asarin_approximate_2000}. On the other hand, generating reachable and invariant sets requires exact sets or inner-approximations. In these approaches, detecting guard set intersections and avoiding explosion in the number of convex sets required to represent the nonconvex reachable set remain key challenges \cite{althoff_set_2021}.

Recent efforts in the literature have brought several new set representations derived for specific applications. These include convex sets with a mix of polyhedral facets and smooth features \cite{silvestreConstrainedConvexGenerators2022}, nonconvex sets closed under polynomial mappings \cite{kochdumperSparsePolynomialZonotopes2021}, and even more expressive functional sets \cite{combastelFunctionalSetsTyped2022a}.
This paper presents a nonconvex set representation named the \textit{hybrid zonotope} that contains both continuous and binary zonotope factors, and is well suited to the analysis of hybrid systems. It is shown that the hybrid zonotope is equivalent to the union of $2^N$ constrained zonotopes through the use of $N$ binary factors. The major contribution of this work is an identity for using hybrid zonotopes to represent exact reachable sets of discrete-time hybrid automata modeled as Mixed Logical Dynamical (MLD) systems \cite{bemporad_control_1999}. 
This identity contains all guard set intersections, changes in dynamics, and reset maps implicitly as properties of the MLD model. 
The resulting reachable set is represented as a single hybrid zonotope equivalent to an exponential number of convex sets while exhibiting linear growth in set representation complexity.
By leveraging binary trees and mixed-integer techniques to identify empty subsets of the hybrid zonotope, it is shown how the complexity growth of the reachable set may be further reduced. 

The remainder of the manuscript is organized as follows. Notation and preliminary information for zonotopes and MLD systems is provided in Section \ref{sec-prelim}. The hybrid zonotope set representation is presented in Section \ref{sec-setrep}. In Section \ref{sec-MLDreach}, a closed-form solution to the forward reachable sets of MLD systems is proven and a redundancy removal technique is described. In Section \ref{sec-binaryTrees}, it is shown how each hybrid zonotope has an associated binary tree that can be used to reduce the number of binary factors required to represent the set. Two numerical examples of the forward reachable sets of MLD systems are provided in Section \ref{sec-numericalEx}. Concluding remarks are given in Section \ref{sec-conclusions}. 
\section{Notation and preliminaries}\label{sec-prelim}

Matrices are denoted by uppercase letters, e.g., $G\in{\rm I\!R}^{n\times n_g}$, and sets by uppercase calligraphic letters, e.g., $\mathcal{Z}\subset{\rm I\!R}^{n}$. Vectors and scalars are denoted by lowercase letters, e.g., $b\in{\rm I\!R}^{n_c}$. 
Commas in subscripts are used to distinguish between properties that are defined for multiple sets, e.g., $n_{g,z}$ describes the complexity of the representation of $\mathcal{Z}$ while $n_{g,w}$ describes the complexity of the representation of $\mathcal{W}$. The $n$-dimensional unit hypercube is denoted by $\mathcal{B}_{\infty}^n=\left\{x\in{\rm I\!R}^{n}~\middle|~\|x\|_{\infty}\leq1\right\}$. The set of all $n$-dimensional binary vectors is denoted by $\{-1,1\}^{n}$. The cardinality of the discrete set $\mathcal{T}$ is denoted by $|\mathcal{T}|$, e.g., $|\mathcal{T}|=8$ for $\mathcal{T}=\{-1,1\}^{3}$.
The concatenation of two column vectors into a single column vector is denoted by $(\xi_1~\xi_2)=[\xi_1^T~\xi_2^T]^T$. The bold $\mathbf{1}$ and $\mathbf{0}$ denote matrices of all $1$ and $0$ elements, respectively, and $\mathbf{I}$ denotes the identity matrix with dimensions indicated by subscripts when not easily deduced from context. 
Given the sets $\mathcal{Z},\:\mathcal{W}\subset{\rm I\!R}^{n},\:\mathcal{Y}\subset{\rm I\!R}^{m}$, and matrix $R\in{\rm I\!R}^{m\times n}$, the linear mapping of $\mathcal{Z}$ by $R$ is $R\mathcal{Z}=\{Rz~|~z\in\mathcal{Z}\}$, the Minkowski sum of $\mathcal{Z}$ and $\mathcal{W}$ is $\mathcal{Z}\oplus\mathcal{W}=\{z+w~|~z\in\mathcal{Z},\:w\in\mathcal{W}\}$, the generalized intersection of $\mathcal{Z}$ and $\mathcal{Y}$ under $R$ is $\mathcal{Z}\cap_R\mathcal{Y}=\{z\in\mathcal{Z}~|~Rz\in\mathcal{Y}\}$, and the union of $\mathcal{Z}$ and $\mathcal{W}$ is $\mathcal{Z}\cup\mathcal{W}=\{x\in{\rm I\!R}^{n}~|~x\in\mathcal{Z}\lor x\in\mathcal{W}\}$. 

\subsection{Zonotopes and constrained zonotopes}

A zonotope is a centrally symmetric, polytopic set representation that is defined as the affine image of a unit hypercube. 
\begin{defn}\label{def-zono} 
    \emph{\cite{mcmullen_zonotopes_1971}} The set $\mathcal{Z}\subset{\rm I\!R}^n$ is a zonotope if there exist $G\in{\rm I\!R}^{n\times n_g}$ and $c\in{\rm I\!R}^{n}$ such that
    \begin{equation}\label{def-eqn-zono}
        \mathcal{Z}=\left\{G\xi+c\:\middle|\:\|\xi\|_\infty\leq1\right\} \: .
    \end{equation}
\end{defn}
The zonotope is given in Generator-representation (G-rep), and the shorthand notation of $\mathcal{Z}=\langle G,c\rangle$ is used to denote the set given by \eqref{def-eqn-zono}. A zonotope is the set of points given by all linear combinations of the center $c$ with the weighted generators---the columns of $G=[g^{(1)}\:\dots\:g^{(n_g)}]$---such that their weights $\xi=(\xi_1\:\dots\:\xi_{n_g})$, called factors, lie within the closed unit hypercube. The complexity of the set is given by the zonotope order $o=n_g/n$. Zonotopes are limited by their symmetry and are not closed under intersection operations \cite{scott_constrained_2016}.
\begin{defn}\label{def-conZono} 
    \emph{\cite{scott_constrained_2016}} The set $\mathcal{Z}_c\subset{\rm I\!R}^n$ is a constrained zonotope if there exist $G\in{\rm I\!R}^{n\times n_g}$, $c\in{\rm I\!R}^{n}$, $A\in{\rm I\!R}^{n_c\times n_g}$, and $b\in{\rm I\!R}^{n_c}$ such that
    \begin{equation}\label{def-eqn-conZono}
        \mathcal{Z}_c=\left\{G\xi+c\:\middle|\:\|\xi\|_\infty\leq1, A\xi=b\right\} \: .
    \end{equation}
\end{defn}
The constrained zonotope is given in Constrained Generator-representation (CG-rep), and the shorthand notation of $\mathcal{Z}_c=\langle G,c,A,b\rangle$ is used to denote the set given by \eqref{def-eqn-conZono}. Through the addition of the linear equality constraints $A\xi=b$ to the projected unit hypercube, the affine image of the constrained space of factors is no longer restricted to be symmetric \cite{scott_constrained_2016}. The degree of freedom order of a constrained zonotope is defined as $o_d=(n_g-n_c)/n$. Constrained zonotopes are closed under linear mappings, Minkowski sums, and generalized intersections.

\subsection{Mixed Logical Dynamical (MLD) systems}

As first introduced in \cite{bemporad_control_1999}, the Mixed Logical Dynamical (MLD) system modeling framework combines continuous and binary variables with logical relations in mixed-integer inequalities to express complex dynamic systems. It has been shown in \cite{bemporad_control_1999,heemels_equivalence_2001} that such a framework can be used to model systems that have mixed continuous and discrete states and inputs, piece-wise affine and bilinear dynamics, finite state machines, qualitative outputs, and those with any combination of the former. An MLD system with linear discrete-time dynamics may be expressed as 
\begin{subequations}\label{eqn-MLDdef}
\begin{align}
x_+=A&x+B_{u}u+B_{w}w+B_{aff}\:,\label{eqn-MLDdef-state}\\
\text{s.t.}\:\:\:E_{x}&x+E_{u}u+E_{w}w\leq E_{aff}\:,\label{eqn-MLDdef-inequality}
\end{align}
\end{subequations}
where $x\in {\rm I\!R}^{n_{xc}}\times\{0,1\}^{n_{xl}}$ are the system states, $u\in {\rm I\!R}^{n_{uc}}\times\{0,1\}^{n_{ul}}$ are the control inputs, and $w \in {\rm I\!R}^{n_{rc}}\times\{0 , 1\}^{n_{rl}}$ are auxiliary variables. The number of inequality constraints is denoted by $n_e$ such that $E_{aff}\in {\rm I\!R}^{n_e}$.

When formulating an MLD model \eqref{eqn-MLDdef}, the so-called ``big-\emph{M}'' constants used in the mixed-integer inequalities to relate continuous values to logical statements are chosen for a user-defined subset of the state space, $\mathcal{X}\subset{\rm I\!R}^{n_{xc}}\times\{0,1\}^{n_{xl}}$, and set of admissible control inputs, $\mathcal{U}\subset{\rm I\!R}^{n_{uc}}\times\{0,1\}^{n_{ul}}$ \cite{bemporad_control_1999}. It follows that for the bounded state-input domain over which the MLD model is defined, the auxiliary variables will belong to a compact set $\mathcal{W}\subset{{\rm I\!R}^{n_{rc}}}\times\{0,1\}^{n_{rl}}$. The MLD representation and set of possible auxiliary variables $\mathcal{W}$ of linear discrete-time hybrid systems may be generated automatically using the modeling tool Hybrid System DEscription Language (HYSDEL) \cite{torrisi_hysdel-tool_2004}.
\section{The hybrid zonotope}\label{sec-setrep}

This section introduces the definition of hybrid zonotopes as an extension of the constrained zonotope through the addition of a vector of binary factors. 
\begin{defn}\label{def-hybridZono}
    The set $\mathcal{Z}_h\subset{\rm I\!R}^n$ is a hybrid zonotope if there exist $G^c\in{\rm I\!R}^{n\times n_{g}}$, $G^b\in{\rm I\!R}^{n\times n_{b}}$, $c\in{\rm I\!R}^{n}$, $A^c\in{\rm I\!R}^{n_{c}\times n_{g}}$, $A^b\in{\rm I\!R}^{n_{c}\times n_{b}}$, and $b\in{\rm I\!R}^{n_c}$ such that
    \begin{equation}\label{def-eqn-hybridZono}
        \mathcal{Z}_h = \left\{ \left[G^c \: G^b\right]\left[\begin{smallmatrix}\xi^c \\ \xi^b \end{smallmatrix}\right]  + c\: \middle| \begin{matrix} \left[\begin{smallmatrix}\xi^c \\ \xi^b \end{smallmatrix}\right]\in \mathcal{B}_\infty^{n_{g}} \times \{-1,1\}^{n_{b}}, \\ \left[A^c \: A^b\right]\left[\begin{smallmatrix}\xi^c \\ \xi^b \end{smallmatrix}\right] = b \end{matrix} \right\}\:.
\end{equation}
\end{defn}

The hybrid zonotope is given in \textit{Hybrid Constrained Generator-representation} (HCG-rep), and the shorthand notation of $\mathcal{Z}_h=\langle G^c,G^b,c,A^c,A^b,b\rangle\subset{\rm I\!R}^n$ is used to denote the set given by \eqref{def-eqn-hybridZono}. When $n_{b}=0$, the hybrid zonotope set representation is equivalent to the constrained zonotope given by Definition \ref{def-conZono}. When $n_{b}\not=0$, the vector of binary factors may take on values from the discrete set ${\{-1,1\}^{n_{b}}}$ containing $2^{n_{b}}$ elements. The degrees of freedom of a hybrid zonotope is a function of both the number of continuous and binary factors with order $o_d=(n_g+n_b-n_c)/n$.
Given that $\|\xi^b\|_{\infty}=1\;\forall\:\xi^b\in\{-1,1\}^{n_{b}}$, the hybrid zonotope is a more general class than the constrained zonotope set representation.
\begin{lem}\label{lem-ZhsubZcsubZ}
\sloppy 
Given any hybrid zonotope 
$\mathcal{Z}_h=\langle G^c,G^b,c,A^c,A^b,b\rangle$,
the zonotope $\mathcal{Z}=\langle [G^c\:G^b],c\rangle$ and constrained zonotope $\mathcal{Z}_c=\langle[G^c\:G^b],c,[A^c\:A^b],b\rangle$ satisfy $\mathcal{Z}_h\subseteq\mathcal{Z}_c\subseteq\mathcal{Z}$.
\vspace{-12 pt}
\begin{pf}
For $\mathcal{Z}$ and $\mathcal{Z}_c$ it holds that $\mathcal{Z}_c\subseteq\mathcal{Z}$ \cite{scott_constrained_2016}. 
For any $z\in\mathcal{Z}_h$ there exist some $\|\xi^c\|_{\infty}\leq1$ and $\xi^b\in\{-1,1\}^{n_b}$ such that $A^c\xi^c+A^b\xi^b=b$ and $z=G^c\xi^c+G^b\xi^b+c$. Letting $\xi=(\xi^c\:\xi^b)$ implies that $\|\xi\|_{\infty}\leq1$, $z=[G^c\:G^b]\xi+c$, and $[A^c\:A^b]\xi=b$, thus $z\in\mathcal{Z}_c$, and therefore $\mathcal{Z}_h\subseteq\mathcal{Z}_c\subseteq\mathcal{Z}$.
\hfill\hfill\qed
\end{pf}
\end{lem}

\vspace{-7 pt}
\subsection{Relation to constrained zonotopes}
\begin{figure}[!htb]
     \centering
     \includegraphics[width=0.85\linewidth]{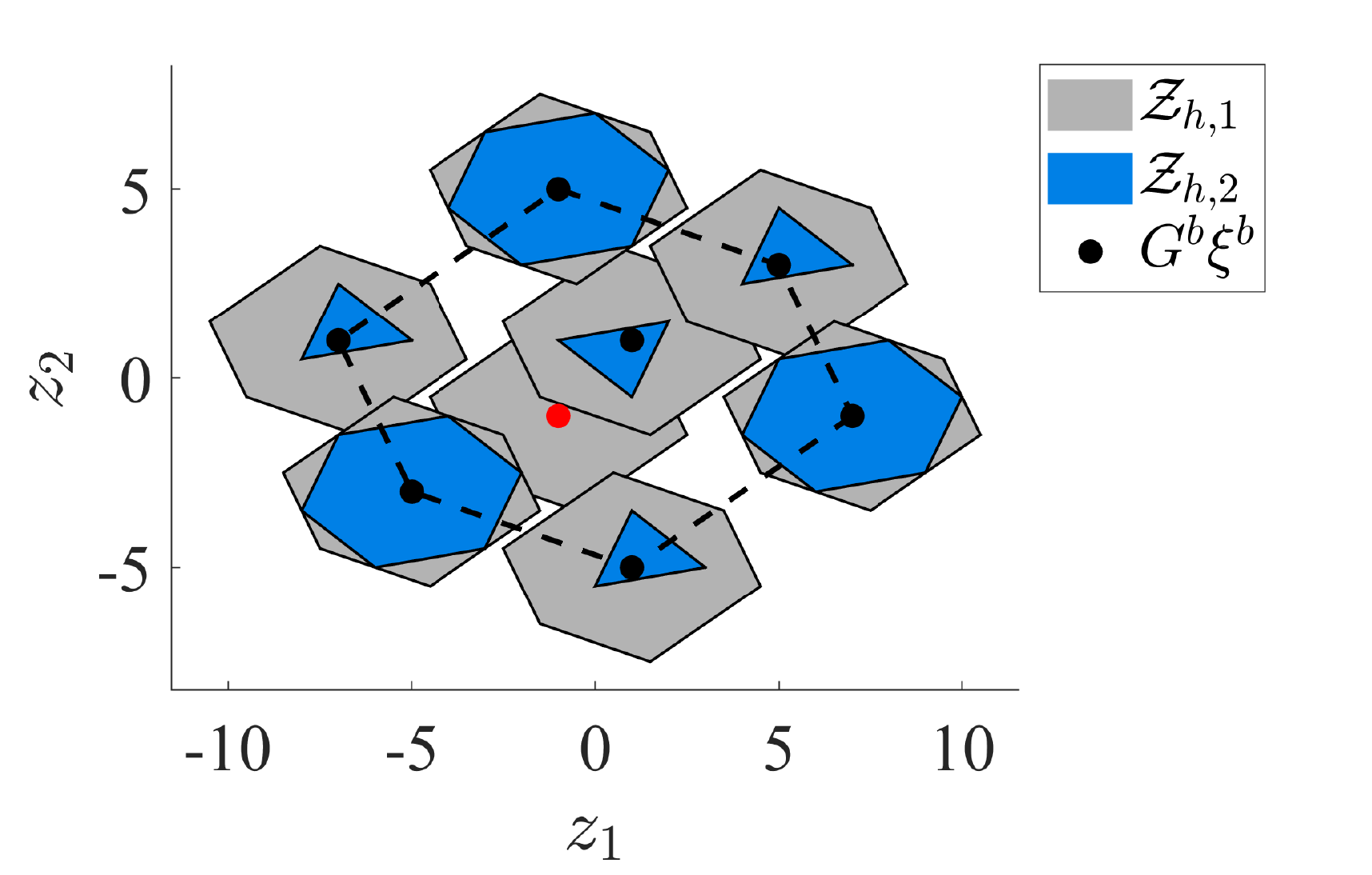}
     \caption{
        Hybrid zonotopes given in Ex. \ref{exa-hybridZono}. Note that the convex hull of the eight discrete points given by $G^b\xi^b$ is equivalent to the zonotope $\mathcal{Z}=\langle G^b,\mathbf{0}\rangle$ as depicted by the dashed lines. The discrete value of the binary factors depicted by the red {\color{red}$\bullet$} results in an infeasible set of continuous constraints of $\mathcal{Z}_{h,2}$ and thus maps to an empty constrained zonotope. }
        \label{fig-Example1}
\end{figure}

The equivalence of the hybrid zonotope with a finite collection of constrained zonotopes is established through the following theorem relying on the closure of hybrid zonotopes under union operations as proven in the ancillary manuscript \cite{birdUnionsComplementsHybrid2021}. 
\begin{thm}\label{thm-ZhIsAUnion}
    The set $\mathcal{Z}_h\subset{\rm I\!R}^n$ is a hybrid zonotope if and only if it is the union of a finite number of constrained zonotopes. 
\vspace{-7 pt}
\begin{pf}
Let $\xi_i^b$ be an entry of the discrete set $\{-1,1\}^{n_{b}}$ containing $2^{n_{b}}$ elements. Define the constrained zonotope
\begin{equation}\label{eqn-thm-decomposedZc}
    \mathcal{Z}_{c,i}=\left\langle G^c,c+G^b\xi^b_i,A^c,b-A^b\xi^b_i\right\rangle\:.
\end{equation}
For any $z\in\mathcal{Z}_{c,i}$ there exists some $\xi^c\in\mathcal{B}_{\infty}^{n_g}$ such that $z=G^c\xi^c+G^b\xi^b_i+c$ and $A^c\xi^c+A^b\xi^b_i=b$. Thus $z\in\mathcal{Z}_h$. Given that the choice of $z$ is arbitrary and the set $\{-1,1\}^{n_{b}}$ is finite,
$\bigcup_{i=1}^{2^{n_{b}}}\mathcal{Z}_{c,i}\subseteq\mathcal{Z}_{h}$.
For any $z\in\mathcal{Z}_h$, there exist some $\xi^c\in\mathcal{B}_{\infty}^{n_g}$ and $\xi^b\in\{-1,1\}^{n_{b}}$ such that $z=G^c\xi^c+G^b\xi^b+c$ and $A^c\xi^c+A^b\xi^b=b$. Also, for $\xi^b=\xi^b_i\Rightarrow z\in\mathcal{Z}_{c,i}$, thus
$\mathcal{Z}_{h}\subseteq\bigcup_{i=1}^{2^{n_{b}}}\mathcal{Z}_{c,i}$
and $\mathcal{Z}_h=\bigcup_{i=1}^{2^{n_{b}}}\mathcal{Z}_{c,i}$. 
Conversely, given any finite collection of constrained zonotopes $\mathcal{Z}_{c,i}\subset{\rm I\!R}^n$ for $i=1,\dots,N$, the hybrid zonotope generated by successive union operations as
$\mathcal{Z}_h=\mathcal{Z}_{c,1}\cup(\mathcal{Z}_{c,2}\cup(\cdots\cup\mathcal{Z}_{c,N}))$,
is an exact representation of the $N$ constrained zonotopes \cite{birdUnionsComplementsHybrid2021}, therefore $\bigcup_{i=1}^{N}\mathcal{Z}_{c,i}=\mathcal{Z}_h$.  \hfill\hfill\qed
\end{pf}
\end{thm}
\vspace{-7 pt}
The hybrid zonotope exhibits the same combinatorial properties as zonotopes, where a symmetric polytope with up to $2\binom{n_g}{n}$ vertices may be represented with $n_g$ continuous factors \cite{mcmullen_zonotopes_1971}. Introducing $n_b$ binary factors, the hybrid zonotope may represent up to $2^{n_b}$ zonotopes. This concept is further explored through the following example.
\begin{exmp}\label{exa-hybridZono}
    Let the set $\mathcal{Z}_c=\left\langle G_z,c_z,A_z,b_z\right\rangle\subset{\rm I\!R}^2$ be the example constrained zonotope given in \emph{\cite{scott_constrained_2016}}, where
    \begin{equation}
        \mathcal{Z}_c=\left\langle\begin{bmatrix}1.5 & -1.5 & 0.5\\1&\phantom{-}0.5&-1\end{bmatrix},\begin{bmatrix}0\\0\end{bmatrix},\begin{bmatrix}1&1&1\end{bmatrix},1\right\rangle \:,\nonumber
    \end{equation}
    and define a hybrid zonotope with continuous generators $G^c=G_z$, binary generators $G^b=2G_z$, and center $c=c_z$ giving $\mathcal{Z}_{h,1}=\left\langle G_z,2G_z,c_z,\emptyset,\emptyset,\emptyset\right\rangle$.
    By adding $n_{b}=3$ binary factors, $\mathcal{Z}_{h,1}$ is equivalent to $2^{n_{b}}=8$ copies of the zonotope $\mathcal{Z}=\left\langle G_z,c_z\right\rangle$ with centers shifted by $2G_z\xi^b\;\forall\:\xi^b\in\{-1,1\}^3$, as depicted in Fig. \ref{fig-Example1}. Including the continuous and binary factors in the equality constraints by defining another hybrid zonotope with $A^c=A^b=A_z$ and $b=b_z$ gives $\mathcal{Z}_{h,2}=\left\langle G_z,2G_z,c_z,A_z,A_z,b_z\right\rangle$,
    as shown in Fig. \ref{fig-Example1}. In contrast to the previous hybrid zonotope, $\mathcal{Z}_{h,2}$ does not represent identical copies. Instead, the linear equality constraints on the continuous factors are also shifted by each of the eight discrete values of the binary factors. When doing so, it is possible that these shifted equality constraints may be infeasible and thus map to empty constrained zonotopes, which happens once in the given example. 
\end{exmp}

The result of Theorem \ref{thm-ZhIsAUnion} provides a method for converting from a hybrid zonotope to a collection of constrained zonotopes, and vice versa, allowing methods developed for the analysis and visualization of other set representations to be applied to hybrid zonotopes. However, the conversion from HCG-rep to a collection of CG-reps, $\mathcal{Z}_{c,i}\:\forall\:i\in\{1,\dots,2^{n_{b}}\}$ given by \eqref{eqn-thm-decomposedZc}, is an enumeration problem that grows exponentially with respect to the number of binary factors. Similarly, the conversion of a collection of constrained zonotopes to a hybrid zonotope may be accomplished algebraically by iteratively applying the union operation derived in \cite{birdUnionsComplementsHybrid2021}, at the cost of increasing the overall complexity of the representation.
Use of the hybrid zonotope is therefore most advantageous when these conversions are not necessary and the representation may be used directly for the analysis of complex dynamical systems, as discussed in the remainder of the manuscript. 
\vspace{-7pt}
\subsection{Basic set operations with hybrid zonotopes}

The identities for linear mappings, Minkowski sums, generalized intersections \cite[Proposition~1]{scott_constrained_2016}, and halfspace intersections \cite[Theorem~1]{raghuramanSetOperationsOrder2022} of constrained zonotopes may be extended to hybrid zonotopes as follows. Beyond these basic set operations, the hybrid zonotope's closure under unions and complements has been proven in the ancillary manuscript \cite{birdUnionsComplementsHybrid2021}.
\begin{prop}\label{prop-setOperations-hybZono}
    \sloppy For any $\mathcal{Z}_h=\langle G_z^c,G_z^b,c_z,A_z^c,A_z^b,b_z\rangle$, $\mathcal{W}_h=\langle G_w^c,G_w^b,c_w,A_w^c,A_w^b,b_w\rangle\subset{\rm I\!R}^n$, $\mathcal{Y}_h=\langle G_y^c,G_y^b,c_y,A_y^c,A_y^b,b_y\rangle\subset{\rm I\!R}^m$, $R\in{\rm I\!R}^{m\times n}$, and $\mathcal{H}^{-}=\{x\in{\rm I\!R}^m~|~l^Tx\leq \rho\}$ the following identities hold:
    \vspace{-20pt}
     {\small
     \begin{align}
        &R\mathcal{Z}_h=\left\langle RG_z^c,RG_z^b,Rc_z,A_z^c,A_z^b,b_z\right\rangle\:, \label{prop-eqn-linMap-hybZono}\\
     &\begin{aligned}
      \mathcal{Z}_h\oplus\mathcal{W}_h=&\left\langle\begin{bmatrix}G_z^c & G_w^c\end{bmatrix},\begin{bmatrix}G_z^b & G_w^b\end{bmatrix},c_z+c_w, \phantom{\begin{bmatrix}b \\b \end{bmatrix}}\right.\\
      &\:\:\:\:\:\left.\begin{bmatrix}A_z^c & \mathbf{0} \\\mathbf{0} & A_w^c\end{bmatrix},\begin{bmatrix}A_z^b & \mathbf{0} \\\mathbf{0} & A_w^b\end{bmatrix},\begin{bmatrix}b_z \\b_w\end{bmatrix} \right\rangle\:, \label{prop-eqn-minkSum-hybZono}
     \end{aligned}\\
    &\begin{aligned}
      &\mathcal{Z}_h\cap_R \mathcal{Y}_h=\left\langle\begin{bmatrix}G_z^c & \mathbf{0}\end{bmatrix},\begin{bmatrix}G_z^b & \mathbf{0}\end{bmatrix},c_z,\phantom{\begin{bmatrix}b \\b \\c\end{bmatrix}}\right.\\
      &\left.\:\:\:\:\:\begin{bmatrix}A_z^c & \mathbf{0} \\\mathbf{0} & A_y^c \\RG_z^c & -G_y^c\end{bmatrix},\begin{bmatrix}A_z^b & \mathbf{0} \\\mathbf{0} & A_y^b \\RG_z^b & -G_y^b\end{bmatrix},\begin{bmatrix}b_z \\b_y \\c_y-Rc_z\end{bmatrix}\right\rangle, \label{prop-eqn-inter-hybZono}
    \end{aligned}\\
     &\begin{aligned}
      &\mathcal{Z}_h\cap_R\mathcal{H}^{-}=\left\langle\begin{bmatrix}G_z^c & \mathbf{0}\end{bmatrix},G_z^b,c_z, \begin{bmatrix}A_z^c & \mathbf{0} \\ l^TRG^c_z & \frac{d_m}{2}\end{bmatrix},\right.\\
      &\:\:\:\:\:\:\:\:\:\:\:\:\:\:\:\:\:\:\:\:\:\:\left.\begin{bmatrix}A_z^b \\ l^TRG^b_z  \end{bmatrix},\begin{bmatrix}b_z \\ \rho-l^TRc_z-\frac{d_m}{2} \end{bmatrix} \right\rangle\:, \\
      & \:d_m=\rho-h^TRc_z+\sum_{i=1}^{n_{g,z}}|h^TRg_z^{(c,i)}|+\sum_{i=1}^{n_{b,z}}|h^TRg_z^{(b,i)}|  \:.\label{prop-eqn-halfSpace-hybZono}
     \end{aligned}
     \end{align}
     }%
\end{prop}
\begin{pf}
The proof follows a straight forward extension of the procedures presented in \cite[Proposition ~1]{scott_constrained_2016} and \cite[Theorem~1]{raghuramanSetOperationsOrder2022} by including the constraint that $\xi^b_i\in\{-1,1\}^{n_{b,i}}$ for $i=z,\:w,\:y$. 
For ease of readability, let $\xi_{i}=(\xi_{i}^c~\xi_{i}^b)$, $G_{i}=[G_{i}^c\:G_{i}^b]$, and $A_{i}=[A_{i}^c~A_{i}^b]$ for $i=z,\:w,\:y$. 

Let $\mathcal{Z}_R$ denote the hybrid zonotope given by the right-hand side of {\eqref{prop-eqn-linMap-hybZono}}. For any point $z\in \mathcal{Z}_h$ there exists some $\xi_z\in\mathcal{B}_{\infty}^{n_{g,z}}\times\{-1,1\}^{n_{b,z}}$ such that $A_z\xi_z=b_z$ and $z=G_z\xi_z+c_z$. Multiplying both sides of $z$ by $R$ gives $Rz=RG_z\xi_z+Rc_z$ and $Rz\in\mathcal{Z}_R$, thus $R\mathcal{Z}_h\subseteq \mathcal{Z}_R$. Conversely, for any point $r\in \mathcal{Z}_R$ there exists some $\xi_r\in\mathcal{B}_{\infty}^{n_{g,z}}\times\{-1,1\}^{n_{b,z}}$ such that $A_z\xi_r=b_z$ and $r=RG_z\xi_r+Rc_z$. Thus there exists some $z\in \mathcal{Z}_h$ such that $Rz=r$. Therefore $\mathcal{Z}_R\subseteq R\mathcal{Z}_h$ and $R\mathcal{Z}_h=\mathcal{Z}_R$.

Let $\mathcal{X}$ denote the hybrid zonotope given by the right-hand side of {\eqref{prop-eqn-minkSum-hybZono}}. For any $z\in \mathcal{Z}_h$ there exists some $\xi_z\in\mathcal{B}_{\infty}^{n_{g,z}}\times\{-1,1\}^{n_{b,z}}$ such that $A_z\xi_z=b_z$ and $z=G_z\xi_z+c_z$. Similarly for any $w\in \mathcal{W}_h$ there exists some $\xi_w\in\mathcal{B}_{\infty}^{n_{g,w}}\times\{-1,1\}^{n_{b,w}}$ such that $A_w\xi_w=b_w$ and $w=G_w\xi_w+c_w$. Let $\xi^c_x=(\xi^c_z~\xi^c_w)$ and $\xi^b_x=(\xi^b_z~\xi^b_w)$. Then $\xi_x\in\mathcal{B}_{\infty}^{n_{g,z}+n_{g,w}}\times\{-1,1\}^{n_{b,z}+n_{b,w}}$ and 
\protect\begin{equation}\label{eqn-minkSum-constraintsX1}
         \begin{bmatrix}A^c_z & \mathbf{0} \\ \mathbf{0} & A^c_w \end{bmatrix}\xi^c_x+\begin{bmatrix}A^b_z & \mathbf{0} \\ \mathbf{0} & A^b_w \end{bmatrix}\xi^b_x=\begin{bmatrix}b^b_z\\b^b_w\end{bmatrix}\: .
\end{equation}
Adding $z$ and $w$ together gives
\protect\begin{equation}\label{eqn-minkSum-zplusw}
    z+w=[G^c_z \: G^c_w]\xi^c_x+[G_z^b \: G_w^b]\xi^b_x+(c_z+c_w)\: ,
\end{equation}
thus $z+w\in\mathcal{X}$ and $\mathcal{Z}_h\oplus\mathcal{W}_h\subseteq \mathcal{X}$. Conversely, for any $x\in\mathcal{X}$ there exists some $\xi_x\in\mathcal{B}_{\infty}^{n_{g,z}+n_{g,w}}\times\{-1,1\}^{n_{b,z}+n_{b,w}}$ such that {\eqref{eqn-minkSum-constraintsX1}} holds and $x=z+w$ as defined by {\eqref{eqn-minkSum-zplusw}}. Letting $\xi^c_x=(\xi^c_z~\xi^c_w)$ and $\xi^b_x=(\xi^b_z~\xi^b_w)$ gives $x\in\mathcal{Z}_h\oplus\mathcal{W}_h$ and $\mathcal{X}\subseteq \mathcal{Z}_h\oplus\mathcal{W}_h$, therefore $\mathcal{Z}_h\oplus\mathcal{W}_h=\mathcal{X}$. 

Let $\mathcal{S}$ denote the hybrid zonotope given by the right-hand side of {\eqref{prop-eqn-inter-hybZono}}. For any $s\in\mathcal{S}$ there exists some $\xi_s\in\mathcal{B}_{\infty}^{n_{g,z}+n_{g,y}}\times\{-1,1\}^{n_{b,z}+n_{b,y}}$ such that
\protect\begin{equation}\label{eqn-minkSum-constraintsX2}
         \begin{bmatrix}A_z^c & \mathbf{0} \\\mathbf{0} & A_y^c \\RG_z^c & -G_y^c\end{bmatrix}\xi^c_s+\begin{bmatrix}A_z^b & \mathbf{0} \\\mathbf{0} & A_y^b \\RG_z^b & -G_y^b\end{bmatrix}\xi^b_s=\begin{bmatrix}b_z \\b_y \\c_y-Rc_z\end{bmatrix}\: ,
\end{equation}
and $s=[G^c_z~\mathbf{0}]\xi^c_s+[G^b_z~\mathbf{0}]\xi^b_s+c_z$. Letting $\xi^c_s=(\xi^c_z~\xi^c_y)$ and $\xi^b_s=(\xi^b_z~\xi^b_y)$ gives $s=G_z\xi_z+c_z$ and $A_z\xi_z=b_z$, thus $s\in \mathcal{Z}_h$. From the final two rows of the equality constraints, $A_y\xi_y=b_y$ and $Rx=G_y\xi_y+c_y$ giving $Rx\in \mathcal{Y}_h$. Therefore $s\in\mathcal{Z}_h\cap_R \mathcal{Y}_h$ and $\mathcal{S}\subseteq \mathcal{Z}_h\cap_R \mathcal{Y}_h$. Conversely, for any $z\in \mathcal{Z}_h\cap_R \mathcal{Y}_h$ there exists some $\xi_z\in\mathcal{B}_{\infty}^{n_{g,z}}\times\{-1,1\}^{n_{b,z}}$ such that $A_z\xi_z=b_z$ and $z=G_z\xi_z+c_z$. Furthermore, there exists some $y\in \mathcal{Y}_h$ such that $y=G_y\xi_y+c_y=Rz$, where $\xi_y\in\mathcal{B}_{\infty}^{n_{g,y}}\times\{-1,1\}^{n_{b,y}}$ and $A_y\xi_y=b_y$. Letting $\xi^c_s=(\xi^c_z~\xi^c_y)$ and $\xi^b_s=(\xi^b_z~\xi^b_y)$ implies that $\xi_s\in\mathcal{B}_{\infty}^{n_{g,z}+n_{g,y}}\times\{-1,1\}^{n_{b,z}+n_{b,y}}$ satisfies {\eqref{eqn-minkSum-constraintsX2}}, and $z=[G^c_z~\mathbf{0}]\xi^c_s+[G^b_z~\mathbf{0}]\xi^b_s+c_z$. Therefore, $z\in\mathcal{S}$, $\mathcal{Z}_h\cap_R \mathcal{Y}_h \subseteq \mathcal{S}$, and $\mathcal{Z}_h\cap_R \mathcal{Y}_h=\mathcal{S}$. 

Let $\mathcal{Q}$ denote the hybrid zonotope given by the right-hand side of {\eqref{prop-eqn-halfSpace-hybZono}}. For any $q\in\mathcal{Q}$ there exists some $\xi_q\in\mathcal{B}_{\infty}^{n_{g,z}+1}\times\{-1,1\}^{n_{b,z}}$ such that 
\protect\begin{equation}\label{eqn-halfInt-constraints}
    \begin{bmatrix}A_z^c & \mathbf{0} \\ h^TRG^c_z & \frac{d_m}{2}\end{bmatrix}\xi^c_q+\begin{bmatrix}A_z^b \\ h^TRG^b_z  \end{bmatrix}\xi^b_q=\begin{bmatrix}b_z \\ f-h^TRc_z-\frac{d_m}{2} \end{bmatrix}\:,
\end{equation}
and $q=[G^c_z~\mathbf{0}]\xi^c_q+G^b_z\xi^b_q+c_z$. Let $\xi_q^c=(\xi_z^c~\xi_h)$ and $\xi_q^b=\xi_z^b$ for $\xi_z^c\in\RN^{n_{g,z}}$, $\xi_h\in\RN$, and $\xi_z^b\in\{-1,1\}^{n_{b,z}}$. Then $q=G_z\xi_z+c_z$ giving $q\in\mathcal{Z}_h$. Expanding the second row of {\eqref{eqn-halfInt-constraints}} gives $h^TR(G_z\xi_z+c_z)=f-(\frac{d_m}{2}\xi_h+\frac{d_m}{2})$. From the definition of $d_m$ and that $\|\xi_h\|_{\infty}\leq1$ it follows that
\protect\begin{equation}
\begin{split}
    h^T &Rq\in\\
    &\left[h^TRc_z-\sum_{i=1}^{n_{g,z}}\vert h^TRg_z^{(c,i)}\vert -\sum_{i=1}^{n_{b,z}}\vert h^TRg_z^{(b,i)}\vert , f \right],
\end{split}
\end{equation}
therefore $Rq\in\mathcal{H}^{-}$ and $\mathcal{Q}\subseteq\mathcal{Z}_h\cap_R\mathcal{H}^{-}$. Conversely, for any point $z\in \mathcal{Z}_h\cap_R\mathcal{H}^{-}$ there exists some $\xi_z\in\mathcal{B}_{\infty}^{n_{g,z}}\times\{-1,1\}^{n_{b,z}}$ such that $A_z\xi_z=b_z$, $z=G_z\xi_z+c_z$, and $h^T R z\leq f$. Thus $h^TRz\in[\alpha,f]$ for some $\alpha\leq h^TRz$ for all $z\in\mathcal{Z}_h\cap_R\mathcal{H}^{-}$. Choose $\alpha=h^TRc_z-\sum_{i=1}^{n_{g,z}}\vert h^TRg_z^{(c,i)}\vert -\sum_{i=1}^{n_{b,z}}\vert h^TRg_z^{(b,i)}\vert$ and let $\beta=h^TRc_z+\sum_{i=1}^{n_{g,z}}\vert h^TRg_z^{(c,i)}\vert +\sum_{i=1}^{n_{b,z}}\vert h^TRg_z^{(b,i)}\vert$, then by Lemma \ref{lem-ZhsubZcsubZ} $h^TR\mathcal{Z}_h\subseteq[\alpha,\beta]$ \cite{leguernicReachabilityAnalysisLinear2010}. Let $\xi_q^c=(\xi_z^c~\xi_{h})$ and $\xi_q^b=\xi_z^b$.
The above then implies that $\xi_q\in\mathcal{B}_{\infty}^{n_{g,z}+1}\times\{-1,1\}^{n_{b,z}}$ satisfies {\eqref{eqn-halfInt-constraints}}, and $z=[G^c_z~\mathbf{0}]\xi^c_q+G^b_z\xi^b_q+c_z\in\mathcal{Q}$. Therefore $\mathcal{Z}_h\cap_R\mathcal{H}^{-}\subseteq\mathcal{Q}$ and $\mathcal{Z}_h\cap_R\mathcal{H}^{-}=\mathcal{Q}$.
\hfill\hfill\qed
\end{pf}

The time complexity of linear mappings given by \eqref{prop-eqn-linMap-hybZono} is $\mathcal{O}(mn(n_g+n_b))$. That of Minkowski sums given by \eqref{prop-eqn-minkSum-hybZono} is $\mathcal{O}(n)$, and that of generalized intersections given by \eqref{prop-eqn-inter-hybZono} is $\mathcal{O}(mn(n_g+n_b))$ and $\mathcal{O}(n)$ when $R=\mathbf{I}_n$. The time complexity of generalized halfspace intersections given by \eqref{prop-eqn-halfSpace-hybZono} is $\mathcal{O}(mn(n_g+n_b))$ and $\mathcal{O}(n(n_g+n_b))$ for $R=\mathbf{I}_n$.
Although performing set operations has low time complexity, the representation complexity of the resulting HCG-rep, i.e., the number of variables and constraints, is increased for all set operations beside linear mappings. 
This increased complexity is then encountered when the resulting set is analyzed. 
Following the evaluation of point containment of constrained zonotopes by solving linear programs \cite[Proposition~2]{scott_constrained_2016}, the hybrid zonotope $\mathcal{Z}_h=\langle G^c,G^b,c,A^c,A^b,b\rangle\subset\RN^{n}$ requires the evaluation of a Mixed-Integer Linear Program (MILP) with constraints
\begin{subequations}\label{eqn-Zh-MILP}
\begin{align}
    &\xi^c\in\mathcal{B}_\infty^{n_{g}}\:,\:\xi^b\in\{-1,1\}^{n_b}\:,\:A^c\xi^c+A^b\xi^b=b\:,\label{eqn-Zh-MILP-constraints}\\
    &z=G^c\xi^c+G^b\xi^b+c\:.\label{eqn-Zh-MILP-generators}
\end{align}
\end{subequations}
Given a point $z\in\RN^n$, if the mixed-integer constraints \eqref{eqn-Zh-MILP-constraints}-\eqref{eqn-Zh-MILP-generators} are feasible then $z\in\mathcal{Z}_h$ by Definition \ref{def-hybridZono}. If the constraints \eqref{eqn-Zh-MILP-constraints} are infeasible then $\mathcal{Z}_h=\emptyset$. 

While solving MILPs to obtain a global optimum is NP-hard, determining their feasibility is NP-complete and may often be decided quickly as compared to performing optimization \cite{achterbergPresolveReductionsMixed2020}. The intersection of a hybrid zonotope and a given halfspace may be detected by determining if $\mathcal{Z}_h\cap\mathcal{H}^{-}=\emptyset$ through \eqref{prop-eqn-halfSpace-hybZono} and evaluating the feasibility of \eqref{eqn-Zh-MILP-constraints}. Alternatively, the bounds of a hybrid zonotope in a direction $l\in{\rm I\!R}^n$ may be found by evaluating the set's support function
\begin{equation}\label{eqn-supportFnZh}
  \rho_{\mathcal{Z}_h}(l)=\max\left\{l^Tz\:\middle|\:z\in\mathcal{Z}_h\right\}\:,
\end{equation}
to generate the supporting halfspace
\begin{equation}\label{eqn-supprtHalfZh}
    \mathcal{H}_l^-=\left\{z\in{\rm I\!R}^{n}\:\middle|\: l^Tz\leq\rho_{\mathcal{Z}_h}(l)\right\}\:.
\end{equation}
When \eqref{eqn-supportFnZh} is solved to obtain a global optimum, the supporting halfspace \eqref{eqn-supprtHalfZh} is tight in the sense that the corresponding hyperplane intersects the set $\mathcal{Z}_h$ and $\mathcal{Z}_h\subset\mathcal{H}_l^-$ \cite{leguernicReachabilityAnalysisHybrid2009}. Note that while the intersection of a hybrid zonotope and a halfspace may be detected either through \eqref{prop-eqn-halfSpace-hybZono} and evaluating the feasibility of \eqref{eqn-Zh-MILP-constraints} or by comparing the support function as $\rho_{\mathcal{Z}_h}(l)\leq \rho$, the former method is less computationally expensive.
\section{Reachable sets of MLD systems}\label{sec-MLDreach}

In this section it is shown how the forward reachable sets of MLD systems can be represented as hybrid zonotopes. It is then shown how the representation complexity of the resulting hybrid zonotope can be reduced by removing redundant equality constraints. 

\subsection{Forward propagation of MLD dynamics}
A closed-form solution to the forward reachable sets of MLD systems as hybrid zonotopes is now presented. 

\begin{thm}\label{thm-MLD-HybridSet}
Consider the MLD system described by \eqref{eqn-MLDdef} with $x \in \mathcal{R}_{k}\subseteq\mathcal{X}\subset{\rm I\!R}^{n_{xc}}\times\{0,1\}^{n_{xl}}$, $u \in \mathcal{U}\subset{\rm I\!R}^{n_{uc}}\times\{0,1\}^{n_{ul}}$, and $w \in \mathcal{W}\subset{\rm I\!R}^{n_{rc}}\times\{0,1\}^{n_{rl}}$ given in HCG-rep. Let
\begin{equation}
    \mathcal{V}=\begin{bmatrix}B_u \\ E_u\end{bmatrix}\mathcal{U}\oplus\begin{bmatrix}B_w \\ E_w\end{bmatrix}\mathcal{W}\oplus\begin{bmatrix}B_{aff}\\\mathbf{0}\end{bmatrix}\:,\nonumber
\end{equation}
and define the polyhedron $\mathcal{H}=\{h\in{\rm I\!R}^{n_e}~|~h\leq E_{aff}\}\subset{\rm I\!R}^{n_{e}}$.
Then the set of states reachable in one time step is given by the hybrid zonotope
\begin{equation}\label{thm-eqn-MLD-HybridSet}
    \mathcal{R}_{k+1}=\left[\mathbf{I}_n~\mathbf{0}\right]\left[\left(\begin{bmatrix}A \\ E_x\end{bmatrix}\mathcal{R}_k\oplus\mathcal{V}\right)\cap_{\left[
  \mathbf{0}~\mathbf{I}_{n_e}\right]}\mathcal{H}\right]\:.
\end{equation}%
\end{thm}%
\begin{pf}
Let $\tilde{\mathcal{R}}$ denote the hybrid zonotope given by the right-hand side of \eqref{thm-eqn-MLD-HybridSet} and $\mathcal{R}_{k+1}$ denote the set of states reachable by the MLD system \eqref{eqn-MLDdef} in one time step. For any $r\in\mathcal{R}_{k+1}$ there exist some $x\in\mathcal{R}_{k}$, $u\in\mathcal{U}$, and $w\in\mathcal{W}$ such that $E_xx+E_uu+E_ww\leq E_{aff}$ and $r=Ax+B_uu+B_ww+B_{aff}$. Let $\Gamma=[A^T~E_x^T]^T$ and
\begin{equation}\label{thm-eqn-mld-hybZ-v}
    v=\begin{bmatrix}B_u \\ E_u\end{bmatrix}u+\begin{bmatrix}B_w \\ E_w\end{bmatrix}w+\begin{bmatrix}B_{aff}\\\mathbf{0}\end{bmatrix}\:.
\end{equation}
Then $v\in\mathcal{V}$ and $(\Gamma x+v)\in\Gamma\mathcal{R}_k\oplus\mathcal{V}$.
Furthermore, $r=\left[\mathbf{I}_n~\mathbf{0}\right](\Gamma x+v)$ and $\left[\mathbf{0}~\mathbf{I}_{n_e}\right](\Gamma x+v)\in\mathcal{H}$. Thus $r\in\tilde{\mathcal{R}}$ and $\mathcal{R}_{k+1}\subseteq\tilde{\mathcal{R}}$.

Conversely, for any $\tilde{r}\in\tilde{\mathcal{R}}$ there exist some $x\in\mathcal{R}_k$ and $v\in\mathcal{V}$ such that $\tilde{r}=\left[\mathbf{I}_n~\mathbf{0}\right]\left(\Gamma x+v\right)$ and $\left[\mathbf{0}~\mathbf{I}_{n_e}\right]\left(\Gamma x+v\right)\in\mathcal{H}$. For any $v\in\mathcal{V}$, there exist some $u\in\mathcal{U}$ and $w\in\mathcal{W}$ such that $v$ is given by \eqref{thm-eqn-mld-hybZ-v}. Then $\tilde{r}=Ax+B_uu+B_ww+B_{aff}$ such that  $E_xx+E_uu+E_ww\in\mathcal{H}\Leftrightarrow E_xx+E_uu+E_ww\leq E_{aff}$. Therefore $\tilde{r}\in\mathcal{R}_{k+1}$, $\tilde{\mathcal{R}}\subseteq\mathcal{R}_{k+1}$, and $\tilde{\mathcal{R}}=\mathcal{R}_{k+1}$.\hfill\hfill\qed
\end{pf}
\begin{rem}
Given that the MLD system \eqref{eqn-MLDdef} is only defined over the bounded subset of the state space $\mathcal{X}$ chosen when formulating the MLD model, the set of states reachable from $\mathcal{R}_k$ in one time step is given by Theorem \ref{thm-MLD-HybridSet} only when $\mathcal{R}_{k}\subseteq\mathcal{X}$.
When applying Theorem \ref{thm-MLD-HybridSet} iteratively to find the set of states reachable for $k=0,\dots,N$ time steps, $\mathcal{R}_N$ may be a subset of the true reachable set if $\mathcal{R}_j\not\subseteq\mathcal{X}$ for some $j\in\{0,\dots,N\}$. This is due to the implicit reduction of the feasible space of the MLD system's mixed-integer inequality constraints caused by introducing big-M constants \cite{lodi_mixed_2010}. The complement of the state space $\mathcal{X}^c$ may be represented as a hybrid zonotope through the methods derived in \cite{birdUnionsComplementsHybrid2021}. The condition that $\mathcal{R}_j\subseteq\mathcal{X}$ may then be verified by determining if $\mathcal{R}_j\cap\mathcal{X}^c=\emptyset$ by Proposition \ref{prop-setOperations-hybZono} and \eqref{eqn-Zh-MILP-constraints}.
\end{rem}

By enforcing the MLD system's mixed-integer inequality constraints as halfspace intersections with hybrid zonotopes, Theorem {\ref{thm-MLD-HybridSet}} provides a method of determining the exact set of states reachable by MLD systems defined by {\eqref{eqn-MLDdef}}. This approach is desirable as the propagation of the system dynamics is given by an identity and is computed algebraically. In contrast with existing approaches \cite{althoff_set_2021,frehse_flowpipe_2013,althoff_computing_2010}, the intersections with guard sets are handled implicitly as properties of the MLD system and require no iterative approximations or optimization programs.
Furthermore, the growth in complexity of the set is a linear function of the number of iterative applications of Theorem \ref{thm-MLD-HybridSet}. Specifically, given an initial set of states $\mathcal{R}_{0}\subset{\rm I\!R}^{n_{xc}}\times\{0,1\}^{n_{xl}}$ and set of admissible control inputs $\mathcal{U}\subset{\rm I\!R}^{n_{uc}}\times\{0,1\}^{n_{ul}}$ in HCG-rep, the set of states reachable by the MLD system \eqref{eqn-MLDdef} in $k$ time steps is a hybrid zonotope $\mathcal{R}_{k}$ with representation complexity 
\begin{subequations}
\begin{align}
    n_{g,r}(k) &=(n_{g,u}+n_{rc}+n_e)k+n_{g,0}\:,\\
    n_{b,r}(k) &=(n_{b,u}+n_{rl})k+n_{b,0}\:,\\
    n_{c,r}(k) &=(n_{c,u}+n_{e})k+n_{c,0}\:.
\end{align}
\end{subequations}
The time complexity of \eqref{thm-eqn-MLD-HybridSet} is dominated by the linear mapping of $\mathcal{R}_0$ and scales as $\mathcal{O}(n(n+n_e)(n_{g,0}+n_{b,0}))$, where $n=n_{xc}+n_{xl}$. Given that $a_1n_{g,0}=a_2n_{b,0}=a_3n_e=n$ for $a_i\in\RN$, the time complexity of $k$ iterations of \eqref{thm-eqn-MLD-HybridSet} scales as $\mathcal{O}(n^3k)$.

\subsection{Redundant inequality constraints}\label{sec-redundantConRemoval}

Each halfspace intersection in {\eqref{thm-eqn-MLD-HybridSet}} introduces an additional ``slack'' factor within the HCG-rep of the resulting reachable set, denoted by $\xi_h\in{\rm I\!R}^{n_{e}}$.
It is possible that some of the inequality constraints of the MLD system \eqref{eqn-MLDdef-inequality} are always satisfied by the elements of $\mathcal{R}_k$ and $\mathcal{U}$ and therefore do not need to be enforced within the hybrid zonotope $\mathcal{R}_{k+1}$. That is, $e_x^ix+e_u^iu+e_w^iw< e_{aff}^i\:\forall\:x\in\mathcal{R}_{k},\:u\in\mathcal{U},$ and $w\in\mathcal{W}$, where $e^i$ is the $i^{th}$ row of the matrix $E$. This redundancy may be detected by evaluating the feasibility of an MILP with constraints
\begin{equation}
\begin{aligned}
    &\left[A_r^c \: A_r^b\right]\left[\begin{smallmatrix}\xi^c_r \\ \xi^b_r \end{smallmatrix}\right] = b \:,\: \xi^b_r\in\{-1,1\}^{n_{b,r}}\\
    &\|(\xi_x^c\:\xi_u^c\:\xi_w^c\:\xi_{h,j\not=i})\|_{\infty}\leq1 \:,\: 1\leq\xi_{h,i}\:,
\end{aligned}
\end{equation}
where $\xi_r^c=(\xi_x^c~\xi_u^c~\xi_w^c~\xi_{h})$, and the $i^{th}$ slack factor, $\xi_{h,i}$, is removed from the infinity norm constraint and instead constrained to be greater than or equal to 1. Note that $\xi_{h,i}\geq-1$ always holds by construction of the halfspace intersection set operation \eqref{prop-eqn-halfSpace-hybZono}.
If the MILP is infeasible, then the $i^{th}$ inequality constraint may be removed thereby reducing the number of constraints and continuous generators in $\mathcal{R}_{k+1}$ by one.
\section{Binary trees}\label{sec-binaryTrees}

In this section it is shown how the enumeration problem of decomposing hybrid zonotopes may be reduced by iteratively growing binary trees in parallel with set operations. It is then shown how the number of binary variables needed to define the hybrid zonotope may be reduced by identifying the nonempty leaves of the binary tree. 

For a hybrid zonotope $\mathcal{Z}_h$, let $\mathcal{T}\subseteq\{-1,1\}^{n_{b}}$  be the set of discrete elements that map to nonempty constrained zonotopes, that is $\mathcal{T}=\{\xi_i^b\in\{-1,1\}^{n_{b}}|\mathcal{Z}_{c,i}\not=\emptyset\}$. Leveraging Theorem \ref{thm-ZhIsAUnion} and $\mathcal{Z}_h\cup\emptyset=\mathcal{Z}_h$, it follows that
$\mathcal{Z}_h=\bigcup_{\xi_i^b\in\mathcal{T}}\mathcal{Z}_{c,i}$.
The enumeration problem in decomposing hybrid zonotopes may therefore be reduced by only considering the values of the binary factors belonging to $\mathcal{T}$. The discrete set $\mathcal{T}$ also gives a measure of how efficient the set is---ideally a hybrid zonotope representing $2^N$ constrained zonotopes would only have $N$ binary factors.

The hybrid zonotope is a mixed-integer set representation and may be described by a rooted binary tree \cite{knuth_art_1997}. The root of the binary tree is the hybrid zonotope $\mathcal{Z}_h$ and the nonempty leaves are the constrained zonotopes $\mathcal{Z}_{c,i}\:\forall\:\xi^b_i\in\mathcal{T}$. The binary tree consists of $n_b$ layers, where the $j^{th}$ layer branches on the value of the $j^{th}$ binary factor. Each layer of the tree between the root and leaves consists of branch nodes given by hybrid zonotopes
\begin{equation}\label{eqn-decompsedZh}
    \mathcal{Z}_{h,i}^{j}=\left\langle G^c,G^b_d,c+G^b_a\xi^b_i,A^c,A^b_d,b-A^b_a\xi^b_i\right\rangle\:,
\end{equation}
where the binary generator and constraint matrices are partitioned such that $G^b=[G^b_a\:G^b_d]$, where $G^b_a$ are the $j$ columns for the ancestor nodes multiplied by $\xi^b_i\in\{-1,1\}^{j}$ for the $i^{th}$ branch node of the layer, and $G^b_d$ the remaining columns for the binary factors that are branched on by the descendants.

\subsection{Growing binary trees in parallel with complex sets}

The set $\mathcal{T}$ may be found with any MILP algorithm that explores the constrained space of factors given by \eqref{eqn-Zh-MILP-constraints}, e.g., branch and cut \cite{lodi_mixed_2010}, and is referred to as the integer feasible set of the MILP.
Although many algorithms exist that may be used to find $\mathcal{T}$, the computational burden grows as the number of variables increases. Through all set operations of hybrid zonotopes, the constraints on the factors of the operating sets are imposed directly in the resulting hybrid zonotope (see Proposition \ref{prop-setOperations-hybZono} and Theorem \ref{thm-MLD-HybridSet}). Thus the hybrid zonotope generated through set operations with additional binary factors may only branch from the nonempty leaves of the operating sets. 

Given a hybrid zonotope $\mathcal{Z}_{h,1}$ with integer feasible set $\mathcal{T}_1\subseteq\{-1,1\}^{n_{b,1}}$, let $\mathcal{Z}_{h,2}$ be a hybrid zonotope found through set operations applied to $\mathcal{Z}_{h,1}$ introducing $k$ additional binary factors. Rather than finding $\mathcal{T}_{2}\subseteq\{-1,1\}^{n_{b,1}+k}$ by solving the MILP \eqref{eqn-Zh-MILP-constraints} for $\mathcal{Z}_{h,2}$ directly, it is possible to leverage the fact that the leaves of $\mathcal{Z}_{h,2}$ are the descendants of $\mathcal{Z}_{h,1}$, where $\mathcal{T}_1$ is already known. Thus an alternative approach is to solve the MILP \eqref{eqn-Zh-MILP-constraints} for the $|\mathcal{T}_1|$ branch nodes given by \eqref{eqn-decompsedZh} at layer $n_{b,1}$, each having only $k$ binary factors. The new integer feasible set $\mathcal{T}_2$ is then given by the union of the results from these $|\mathcal{T}_1|$ MILPs appended to the values of $\mathcal{T}_1$. This approach is described in Algorithm \ref{alg-growBinaryTree}. 

\begin{algorithm}[!ht]
\caption{Branching the binary tree of $\mathcal{Z}_{h,2}$ on the descendants of $\mathcal{Z}_{h,1}$.}\label{alg-growBinaryTree}
\algorithmicrequire{ $\mathcal{Z}_{h,2}=\langle G^c,G^b,c,A^c,A^b,b\rangle$, $\mathcal{T}_{1}\subseteq\{-1,1\}^{n_{b,1}}$}\\
\algorithmicensure{ $\mathcal{T}_{2}\subseteq\{-1,1\}^{n_{b,2}}$}
\begin{algorithmic}[1]
\For {$\xi^b_i\in\mathcal{T}_1$}
\State $\mathcal{Z}^{n_{b,1}}_{h,i}\gets$ \eqref{eqn-decompsedZh} for $\xi^b_i$
\State Solve MILP to find integer feasible set $\mathcal{T}$ of $\mathcal{Z}^{n_{b,1}}_{h,i}$
\State Append entries of $\mathcal{T}$ to $\xi^b_i$ and store in $\mathcal{T}_2$
\EndFor
\end{algorithmic}
\end{algorithm}

Since finding $\mathcal{T}$ amounts to an exhaustive search of the integer feasible space of the MILP \eqref{eqn-Zh-MILP-constraints}, Algorithm \ref{alg-growBinaryTree} aims to reduce the number of branches that must be searched at each iteration by solving more, smaller MILPs. Each of these smaller MILPs searches the subtrees branching on the binary factors added since the last search has been performed. Thus leveraging information stored in the set $\mathcal{T}_1$ prevents searching nodes that have already been determined as infeasible during previous iterations. Note that Algorithm \ref{alg-growBinaryTree} is NP-hard with worst-case exponential run time. Nevertheless, this approach may allow the decomposition of complex hybrid zonotopes into a collection of constrained zonotopes when many set operations are applied iteratively.

\subsection{Reducing the number of binary factors}\label{sec-redundantBinRemoval}

Given a hybrid zonotope with $n_b>\log_2(\mathcal{|T|})$, it is possible that the set may be represented with a reduced number of binary factors.
Once $\mathcal{T}$ is known, linearly dependent binary factors may be detected and removed as follows. First, let $T\in{\rm I\!R}^{n_{b}\times|\mathcal{T}|}$ be a matrix with each column an element of $\mathcal{T}$, thus $T(i,j)=\pm1\:\forall\:i,j$. Let $n_\phi=\text{rank}(T)$, if $n_\phi<n_{b}$ then there exists a linear mapping
$M_1T(\Phi,\cdot)=T$,
where $\Phi\in\mathbb{N}^{n_\phi}_+$ are the indices of the linearly independent rows of $T$. Thus the hybrid zonotope $\mathcal{Z}_h=\langle G^c,G^b,c,A^c,A^b,b\rangle$ is equivalent to $\mathcal{Z}_h^r=\langle G^c,G^bM_1,c,A^c,A^bM_1,b\rangle$ where $\mathcal{Z}_h^r$ has $n_{b}^{r}=n_\phi<n_{b}$ binary factors. The integer feasible set of $\mathcal{Z}_h^r$ is then given by
$\mathcal{T}^r=\bigcup_{i=1}^{n_{\phi}} T(\Phi,i)$.

If all feasible values of a binary factor are the same then it can be removed as follows. Let $T(\Phi)$ be sorted such that the constant linearly independent row occurs first, i.e., $T(\Phi(1),\cdot)=1$ or $-1$, and let
\begin{equation}
    M_2=M_1\begin{bmatrix}\mathbf{0}_{1\times n_\phi-1}\\\mathbf{I}_{n_\phi-1}\end{bmatrix}\:,\: m_2=M_1\begin{bmatrix}T(\Phi(1),1)\\\mathbf{0}_{n_\phi-1\times 1}\end{bmatrix}\:.
\end{equation}
Then $\mathcal{Z}_h=\langle G^c,G^b,c,A^c,A^b,b\rangle$ is equivalent to $\mathcal{Z}_h^r=\langle G^c,G^bM_2,c+G^bm_2,A^c,A^bM_2,b-A^bm_2\rangle$ where $\mathcal{Z}_h^r$ has $n_{b}^{r}=n_\phi-1<n_{b}$ binary factors. The integer feasible set of $\mathcal{Z}_h^r$ is then given by
$\mathcal{T}^r=\bigcup_{i=1}^{n_{\phi}} T(\Phi_2,i)$,
where $\Phi_2=\Phi(j)$ for $j=2,\dots,n_\phi$. 

The linearly independent columns of the matrix $T$ may be found through QR decomposition, and the matrix $M_1$ may be found using the Moore–Penrose inverse. The total time complexity of these operations scales as $\mathcal{O}(n_b|\mathcal{T}|^2)$. Although the number of binary factors, and equivalently the number of layers in the binary tree, are reduced, the nonempty leaves of the binary tree are not changed \cite{knuth_art_1997}. Thus detecting and removing redundancy in the binary factors through the described approach reduces the complexity of the hybrid zonotope set representation without altering the set. To further reduce computational complexity, future work on hybrid zonotopes will focus on joint order reduction
to provide inner- and outer-approximations.
\begin{rem}
The proposed method of removing redundant binary variables is an application of aggregating implied free variables within MILPs. The method described here is rigorous and exact; however, approximations may be used as done during the presolve stage of commercial MILP solvers \cite{achterbergPresolveReductionsMixed2020}.
\end{rem}
\section{Numerical examples}\label{sec-numericalEx}

This section presents the forward reachable sets of two MLD systems in the form of \eqref{eqn-MLDdef}.
MLD representations of the presented hybrid systems are obtained using HYSDEL 3.0 \cite{torrisi_hysdel-tool_2004}. Optimization problems are solved using Gurobi \cite{gurobi_optimization_gurobi_2021}. Figures are generated by decomposing the hybrid zonotope into a collection of constrained zonotopes by Theorem \ref{thm-ZhIsAUnion} and converting them to H-rep polytopes. If the order of the constrained zonotopes are below 100, they are converted to an H-rep polytope using the Multi-Parametric Toolbox (MPT) \cite{herceg_multi-parametric_2013}, otherwise tight over-approximations are found by sampling the support function \eqref{eqn-supportFnZh} in 250 uniformly-distributed directions.
Numerical results are generated with MATLAB on a desktop computer using 4 cores of a 3.0 GHz Intel i7 processor with 32 GB of RAM. 

\subsection{Piece-wise affine system with two equilibrium points}\label{sec-numericalEx-shiftedEq}
\begin{figure}[!htb]
     \centering
     \includegraphics[width=0.85\linewidth]{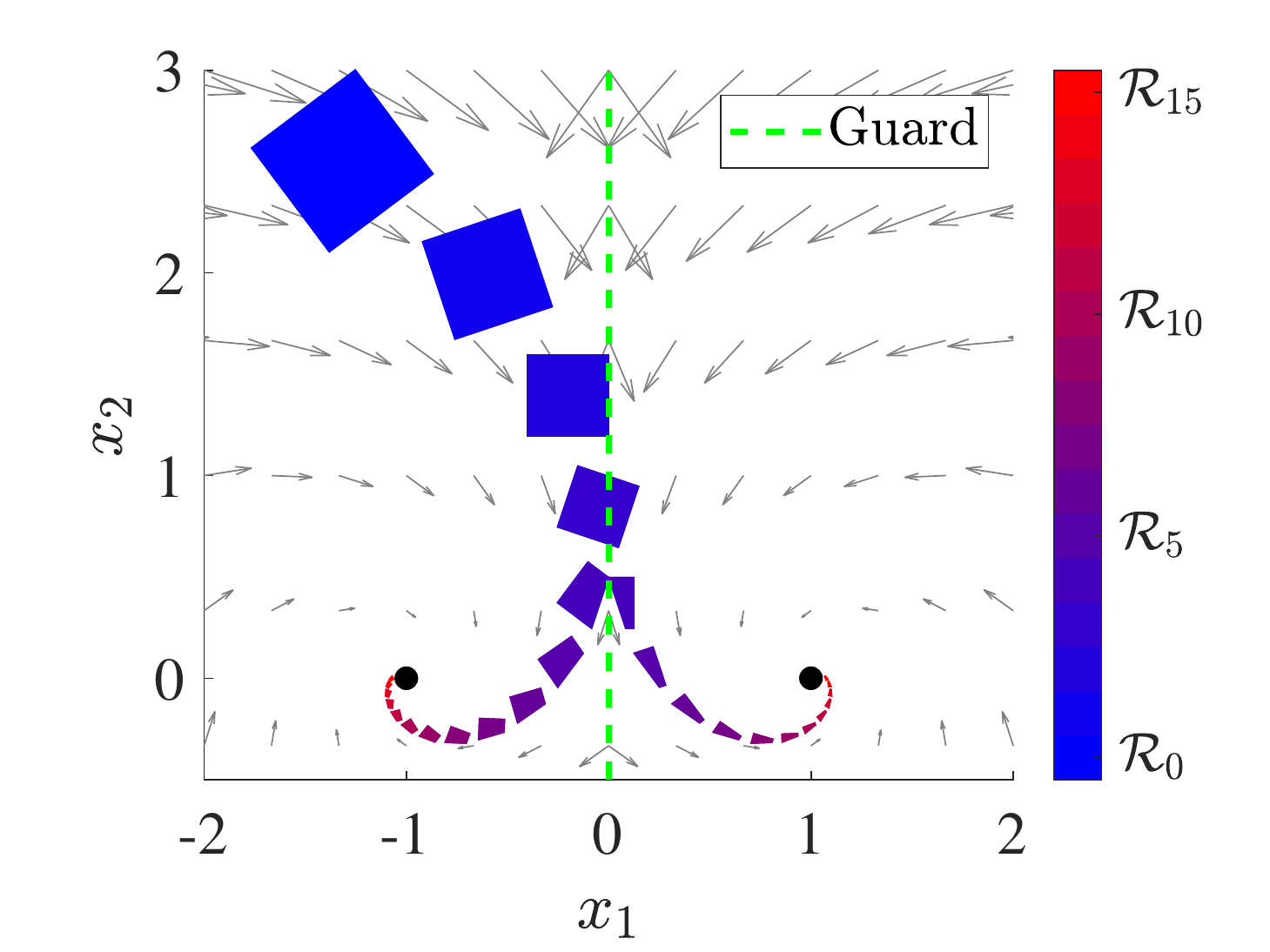}
     \caption{Reachable set of PWA system \eqref{eqn-twoEq-PWA} with two subsystems, each having an equilibrium point depicted by $\bullet$ and autonomous dynamics with vector fields depicted by $\rightarrow$.}
    \label{fig-Example2}
\end{figure}

Consider the discrete-time Piece-Wise Affine (PWA) system given by
\begin{equation}\label{eqn-twoEq-PWA}
  x[k+1]=\begin{cases}
            \begin{bmatrix}
              \hphantom{-}0.75 & 0.25 \\
              -0.25& 0.75
            \end{bmatrix}x[k]+\begin{bmatrix}
                             -0.25 \\
                             -0.25
                           \end{bmatrix}, & \mbox{if} \:x_1\leq0\:, \\
            \begin{bmatrix}
              0.75 & -0.25 \\
              0.25 & \hphantom{-}0.75
            \end{bmatrix}x[k]+\begin{bmatrix}
                             \hphantom{-}0.25 \\
                             -0.25
                           \end{bmatrix}, & \mbox{otherwise}\:.
          \end{cases}
\end{equation}
This hybrid system consists of two stable, autonomous subsystems, each having an equilibrium point at $x=\pm[1~0]^T$. The PWA system can be represented as an MLD system by introducing two continuous auxiliary variables, $n_{rc}=2$, one binary auxiliary variable, $n_{rl}=1$, and ten inequality constraints, $n_e=10$. The states reachable by \eqref{eqn-twoEq-PWA} in $N=15$ time steps
from the initial set given by the zonotope 
\begin{equation}
    \mathcal{R}_0=\left\langle \begin{bmatrix}
        0.25 & -0.19 \\
        0.19 & 0.25
    \end{bmatrix},\begin{bmatrix}
        -1.31 \\ 2.55
    \end{bmatrix} \right\rangle\:,\nonumber
\end{equation}%
 are shown in Fig. \ref{fig-Example2}.
 The set representation dimensions and computation times are given in Table \ref{tab-example2Eqs} for the reachable sets with and without redundancy removal.

\begin{table}[!htb]
\caption{Results of reachability analysis for \eqref{eqn-twoEq-PWA} with redundancy removal, $\mathcal{R}_{15}^r$, and without, $\mathcal{R}_{15}$. Reported computation times include all steps from initializing $\mathcal{R}_0$ to generating the set with shown dimensions.}
\begin{center}
  \begin{tabular}{l c c c c c}
  \hline
  Set & $n_{g,r}$ & $n_{b,r}$ & $n_{c,r}$ & $|\mathcal{T}|$ & Time (s) \\
  \hline
  $\mathcal{R}_{15}$ & 182 & 15 & 150 & 2 & 0.02 \\
  $\mathcal{R}_{15}^r$ & 142 & 1 & 110 & 2 & 0.36 \\
  \hline
  \end{tabular}
\end{center}
\label{tab-example2Eqs}
\end{table}

The auxiliary binary variable in the MLD representation of this PWA system indicates on which side of the guard a state is located. If the reachable set is fully contained on one side of the guard, then only one value of the binary auxiliary variable is feasible. When a guard crossing occurs at $k=3$, the introduced binary variable has a feasible value of $-1\lor 1$ and branches the reachable set. No additional guard crossings occur in all following time steps. Thus the feasible value of the subsequent binary variables are dependent on that of the one introduced at $k=3$---i.e., a state can only be on the right-hand side of the guard if it is a trajectory from a state crossing the guard at $k=3$. The redundancy removal techniques given in Section \ref{sec-redundantBinRemoval} capture these dependencies to reduce the reachable set having fifteen binary factors and a full binary tree with $2^{15}=32,768$ leaves to one with \emph{two leaves from a single binary factor}. In this example, $40$ of the inequality constraints are identified as redundant and removed using the method described in Section \ref{sec-redundantConRemoval}.

\subsection{Thermostat-controlled heated rooms}\label{sec-numericalEx-heatedRooms}
This example extends the heated room scenario given in \cite{althoff_computing_2010}, where the thermostatic control and heat exchange among adjacent rooms is modeled as a hybrid system. The continuous temperature dynamic of the $i^{th}$ room is modeled as 
\begin{equation}\label{eqn-heatedRooms-CT}
    \dot{x}_i=c\cdot h_i+b_i(u-x_i)+\sum_{i\not=j}a_{ij}(x_j-x_i)\:,
\end{equation}
where the heat transfer coefficient $a_{ij}$ is $ 1 $ between adjacent rooms and $0$ otherwise, the heat transfer coefficient between the room and the outside environment is $b_i=0.08q$ where $q$ is the number of exposed walls, the heating power is $c=15$ with $h_i\in\{0,1\}$ for rooms with heaters and $h_i=0$ otherwise, and the outside temperature may take on any value within the interval $u\in[0,0.1]$ \cite{althoff_computing_2010}. Heaters located in select rooms are controlled by discrete-time thermostats that turn on when the sampled temperature in the room decreases below $22^{\circ}$C and turn off when it increases above $24^{\circ}$C. The closed-loop temperature dynamics of the building may be modeled as an MLD system by introducing one binary state, three auxiliary binary variables, and nine inequality constraints for each heater. 

\begin{figure}[!htb]
     \centering
     \includegraphics[width=0.75\linewidth]{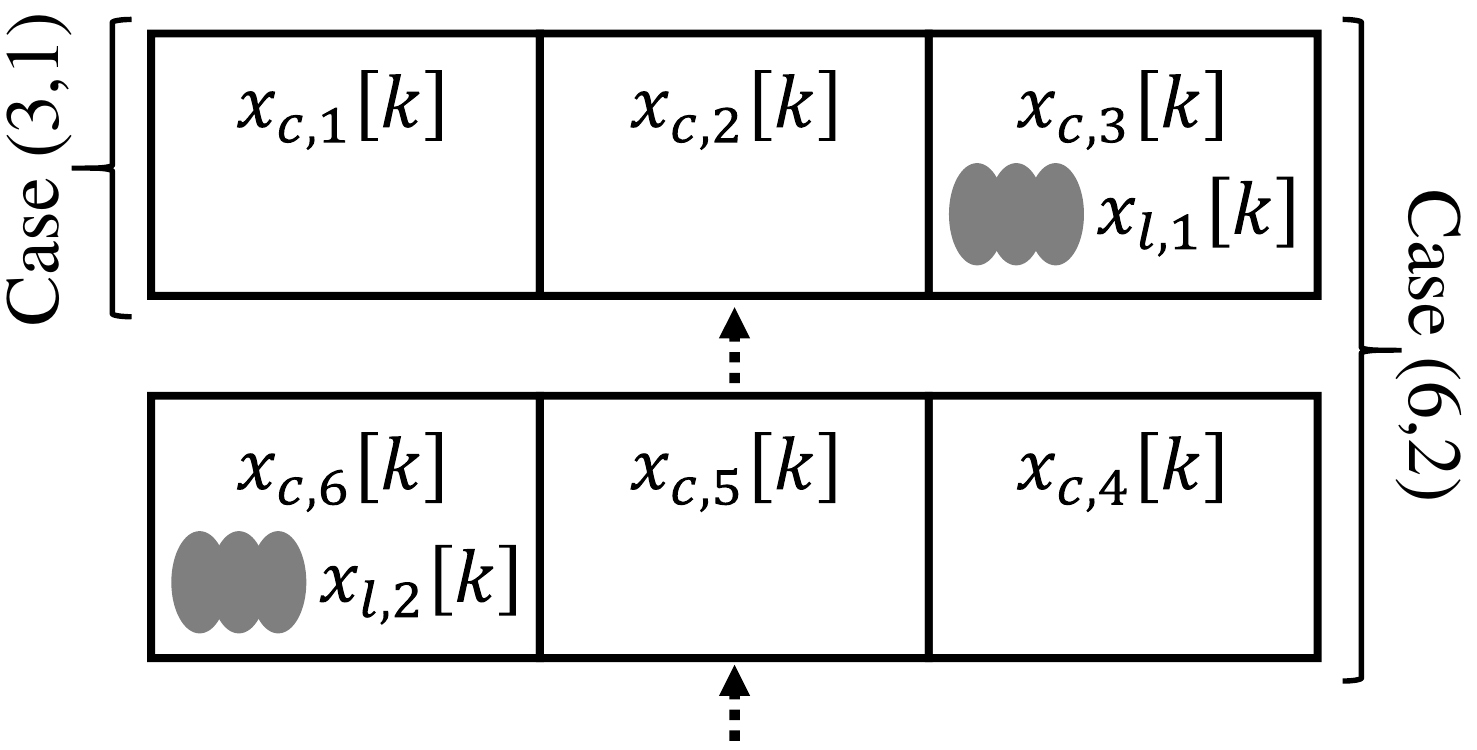}
     \caption{
     Room layout and heater locations for a varying number of rooms. The pattern shown is repeated for Case$(9,3)$ and Case$(12,4)$.}
    \label{fig-HeatedRooms_Diagram}
\end{figure}

\begin{figure*}[!htb]
     \centering
     \includegraphics[width=0.9\linewidth]{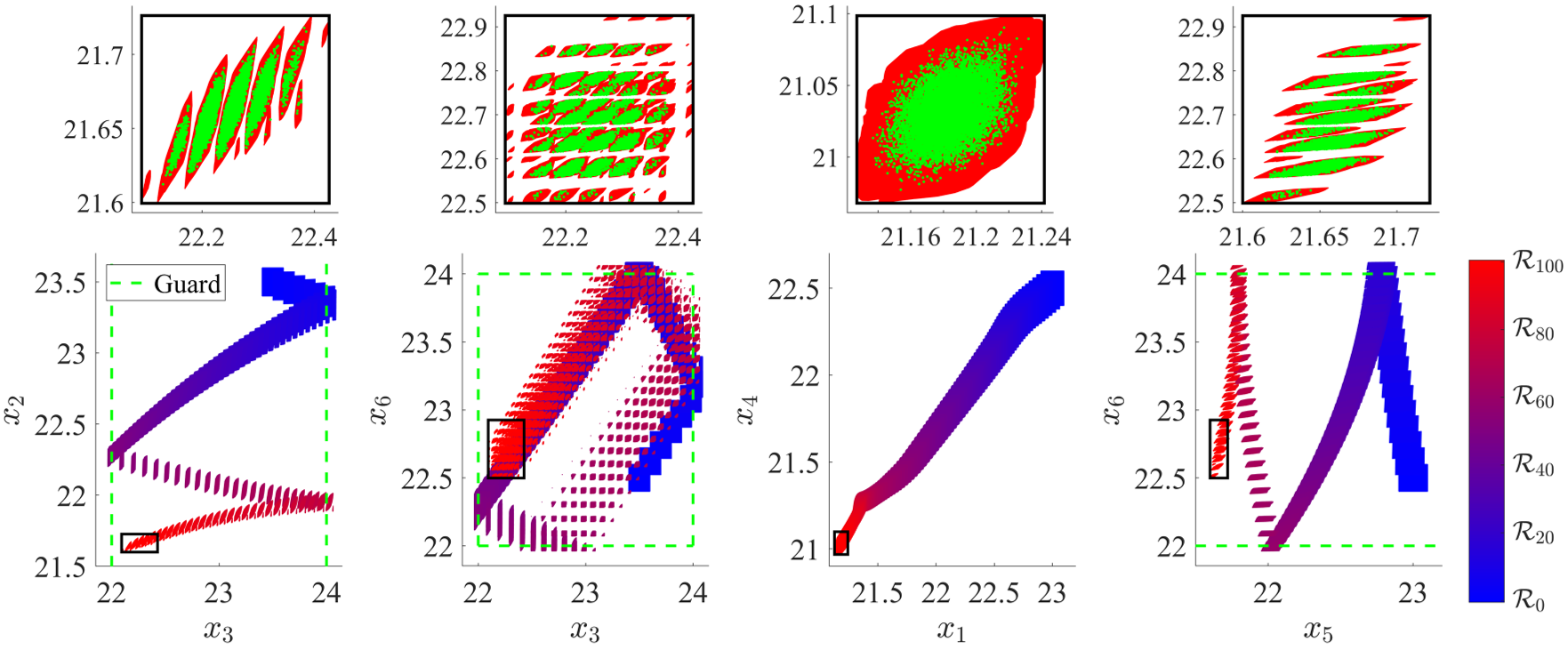}
     \caption{Projections of the reachable set of the heated room MLD system Case$(6,2)$. Supporting halfspaces in each state dimension at the final time step shown by black boxes. Zoomed in plots of the final reachable set shown with $10^4$ randomly sampled, simulated trajectories given by green dots. 
     }
    \label{fig-Example_HeatedRoom_subp_Ts001}
\end{figure*}

Four cases are considered with $n_{xc}=3p$ continuous and $n_{xl}=1p$ discrete states for $p=1,\dots,4$. Each case is coded as Case$(n_{xc},n_{xl})$ to denote the varying building layout shown in Fig. \ref{fig-HeatedRooms_Diagram}. Heaters are located in every third room such that $h_j\in\{0,1\}$ for $j=3p$ as depicted in Fig. \ref{fig-HeatedRooms_Diagram}. Using a discrete time step of $T_s=0.01$ and a zero-order-hold discrete transform of the continuous dynamics \eqref{eqn-heatedRooms-CT}, the reachable set of the four MLD systems for a time interval of $t=[0,1]$ is generated as hybrid zonotopes with dimensions reported in Table \ref{tab-HeatedRooms_HCGdims}. 
The set of initial states of Case$(n_{xc},n_{xl})$ is given by $\mathcal{R}_0=[\mathbf{I}_{n_{xc}}~\mathbf{0}]\mathcal{X}_0\times\{1\}^{n_{xb}}$, where $\mathcal{X}_0=(s~s)\oplus[-0.1,0.1]^{12}$ for $s=[23~23.5~23.5~22.5~23~22.5]^T$.
The computation time of each step of the proposed method is provided in Table \ref{tab-HeatedRooms_HCGtimes}. Four 2D projections of the reachable set for Case$(6,2)$ are plotted in Fig. \ref{fig-Example_HeatedRoom_subp_Ts001}.

In Table \ref{tab-HeatedRooms_HCGdims} it is shown how hybrid zonotopes are able to capture possible exponential growth in the complexity of the nonconvex reachable set with linear growth in set representation complexity. In Case$(3,1)$, the hybrid zonotope $\mathcal{R}_{100}$ is equivalent to the union of 39 convex subsets using 1003 continuous and 300 binary factors. As the complexity of the system is increased through the other three cases, more guard crossings occur over the $100$ step horizon and the resulting reachable set is increasingly nonconvex. In Case$(12,4)$, the hybrid zonotope $\mathcal{R}_{100}$ is equivalent to the union of over $410\times10^3$ convex subsets using 3712 continuous and 1200 binary factors. After applying the proposed redundancy removal techniques, the reduced hybrid zonotope $\mathcal{R}_{100}^r$ represents the nonconvex reachable set using only $484$ continuous factors, $58$ binary factors, and $372$ constraints.

\begin{table}[!htb]
\caption{Set dimensions of reachability analysis for the heated rooms with redundancy removal, $\mathcal{R}_{100}^r$, and without, $\mathcal{R}_{100}$.}
\setlength\tabcolsep{5pt}
\begin{center}
  \begin{tabular}{c c c c c c c c}
  \hline
  & \multicolumn{3}{c}{$\mathcal{R}_{100}$} & \multicolumn{3}{c}{$\mathcal{R}_{100}^r$} \\
  \cmidrule(lr){2-4}\cmidrule(lr){5-7}
   Case & $n_{g,r}$ & $n_{b,r}$ & $n_{c,r}$ & $n_{g,r}$ & $n_{b,r}$ & $n_{c,r}$ & $|\mathcal{T}|$\\
  \hline
  $(3,1)$ & 1003 & 300 & 900 & 261 & 19 & 113 & 39 \\
  $(6,2)$ & 1906 & 600 & 1800 & 283 & 29 & 177 & 657\\
  $(9,3)$ & 2809 & 900 & 2700 & 445 & 64 & 336 & 66523 \\
  $(12,4)$ & 3712 & 1200 & 3600 & 484 & 58 & 372 & 410605 \\
  \hline
  \end{tabular}
\end{center}
\label{tab-HeatedRooms_HCGdims}
\end{table}

The scalability of the proposed approach can be seen in Table \ref{tab-HeatedRooms_HCGtimes}. The computational complexity of Theorem \ref{thm-MLD-HybridSet} to find the unreduced set $\mathcal{R}_{100}$ scales as $\mathcal{O}(n^3)$ and is reflected in the reported computation times. On the other hand, the use of Algorithm \ref{alg-growBinaryTree} to explore the hybrid zonotope's binary tree is NP-hard. However, the complexity of the binary tree is a direct consequence of the number of discrete changes in the hybrid dynamics of the system. In Case$(3,1)$ the number of nonempty leaves of the binary tree is relatively small, and Algorithm \ref{alg-growBinaryTree} contributes only $4\%$ of the total computation time. In Case$(12,4)$, the time spent on Algorithm \ref{alg-growBinaryTree} jumps to nearly $92\%$ of the total computation time. However when comparing this value to $|\mathcal{T}|$, the average time spent per nonempty leaf explored only ranges from $2.2-7.3$ms across all cases. The computational burden of detecting redundant inequality constraints grows with the representation complexity of the hybrid zonotope; however, the number of evaluations of the NP-complete problem is a function of the number of constraints.
\begin{table}[!htb]
\caption{Computation times in seconds for all operations within the reachability analysis. The total time to find the reduced set $\mathcal{R}_{100}^r$ is given by the sum of the individual operations.}
\begin{center}
  \begin{tabular}{c c c c c c}
  \hline
  Case & Theorem \ref{thm-MLD-HybridSet} & Alg. \ref{alg-growBinaryTree} & Redundancy & Total \\
  \hline
  $(3,1)$ & 0.20 & 0.11 & 2.17 & 2.49 \\
  $(6,2)$ & 0.84 & 1.43 & 8.15 & 10.41 \\
  $(9,3)$ & 2.24 & 168.91 & 51.33 & 222.49 \\
  $(12,4)$ & 3.87 & 3001.54 & 264.30 & 3269.71 \\
  \hline
  \end{tabular}
\end{center}
\label{tab-HeatedRooms_HCGtimes}
\end{table}

\vspace{-7 pt}
Tables \ref{tab-M1Times} and \ref{tab-M2Times} compare the use of hybrid zonotopes to represent the reachable set of the thermostat-controlled heated rooms to two existing exact methods:
\begin{enumerate}
    \item[M1] represent the reachable set as a collection of constrained zonotopes generated using the algorithm described in \cite[Algorithm~1]{leguernicReachabilityAnalysisHybrid2009},\label{comp-collectionZono}
    \item[M2] represent the reachable set by iterating over the mixed-integer constraints of the MLD system given by \eqref{eqn-MLDdef} as described in \cite{bemporad_verification_1999} and implemented using YALMIP \cite{lofberg_yalmip_2004}.\label{comp-MLDimp} 
\end{enumerate}
While both of these methods provide the same reachable set as the proposed method using hybrid zonotopes, there are distinctions. Computing reachable sets using M1 results in a worst-case exponential growth in representation complexity, as shown in Table \ref{tab-M1Times}. Furthermore, the resulting reachable set consists of multiple convex sets, thus requiring each convex set to be analyzed to verify properties of the nonconvex reachable set; e.g., in Case$(9,3)$ a total of 13590 linear programs would need to be solved for each guard to detect crossings at time step $k=101$. This growth in complexity resulted in the final Case$(12,4)$ being terminated on the $86^{th}$ time step after 20 hours of computation time. Note that M1 generates fewer convex sets when compared to decomposing the hybrid zonotopes given by Theorem \ref{thm-MLD-HybridSet}. This is because M1 only detects guard crossings that cause the heater to change states, where Theorem \ref{thm-MLD-HybridSet} branches along all guards in the MLD representation.
The reachable set given by M2, on the other hand, is compact and fast to generate, as shown in Table \ref{tab-M2Times}. Similar to the results of Theorem \ref{thm-MLD-HybridSet}, the use of the MLD system model results in only linear growth in representation complexity. The proposed approach and M2 formulate the same feasible space of a MILP in different ways, and are complementary methods. The advantage of hybrid zonotopes is that they lend themselves to use in other set-theoretic methods leveraging the set operations in Proposition \ref{prop-setOperations-hybZono}.

\begin{table}[!htb]
\caption{Computation time and representation complexity for method M1. The total number of generators and constraints summed over the collection of constrained zonotopes is reported by $n_g$ and $n_c$ respectively. Analysis that did not finish are denoted by DNF.}
\begin{center}
  \begin{tabular}{c | c c c c }
  \hline
   Case & $n_{g}$& $n_{c}$ & Sets & Time (s) \\
  \hline
  $(3,1)$ & 1833 & 82 & 17 & 7.99 \\
  $(6,2)$ & 32807 & 2597 & 285 & 118.95 \\
  $(9,3)$ & 1710990 & 229680 & 13590 & 10015.69 \\
  $(12,4)$ & DNF & DNF & DNF & $>72\times10^{3}$\\
  \hline
  \end{tabular}
\end{center}
\label{tab-M1Times}
\end{table}

\begin{table}[!htb]
\caption{Computation time and representation complexity for method M2. The number of continuous variables is given by $n_g$, the number of binary variable is given by $n_b$, and the number of \emph{inequality} constraints is given by $n_c$. }
\begin{center}
  \begin{tabular}{c | c c c c }
  \hline
   Case & $n_{g}$ & $n_b$ & $n_{c}$ & Time (s) \\
  \hline
  $(3,1)$ & 403 & 401 & 2315 & 0.71 \\
  $(6,2)$ & 706 & 802 & 4430 & 0.72 \\
  $(9,3)$ & 1009 & 1203 & 6545 & 0.71 \\
  $(12,4)$ & 1312 & 1604 & 8660 & 0.74 \\
  \hline
  \end{tabular}
\end{center}
\label{tab-M2Times}
\end{table}

\begin{figure}[!htb]
     \centering
     \includegraphics[width=0.82\linewidth]{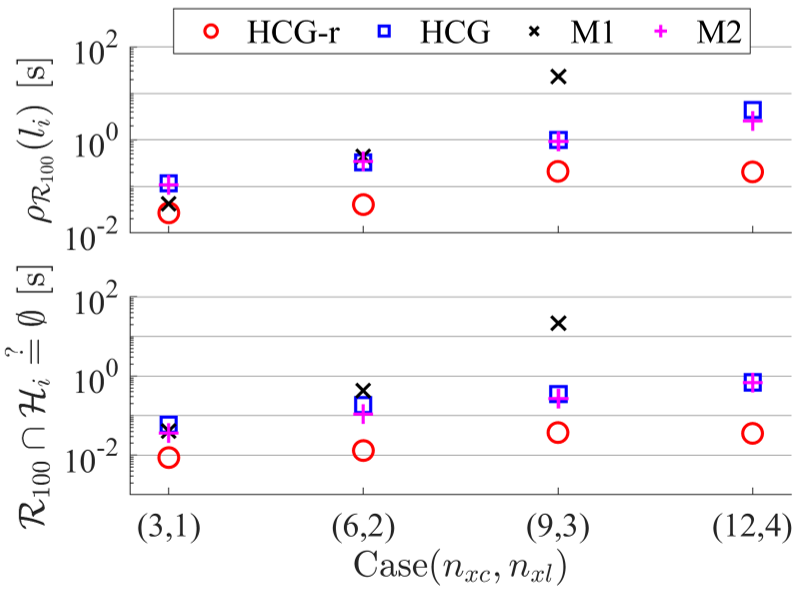}
     \caption{
     Average time to compute support functions and to detect halfspace intersections with the reachable sets. }
    \label{fig-HalfandSupTimes}
\end{figure}

The time to detect halfspace intersections and evaluate support functions \eqref{eqn-supportFnZh} using hybrid zonotopes with and without redundancy removal, denoted by HCG-r and HCG respectively, is compared to the same operations using methods M1 and M2 in Fig. \ref{fig-HalfandSupTimes}. In these results, the direction vector $l_i\in\RN^{n_{xc}}$ is randomly sampled 100 times and the support function $\rho_{\mathcal{R}_{100}}(l_i)$ solved to find a global optimum. To provide a realistic comparison of halfspace detection, the 100 halfspaces are split into 50 true---$\mathcal{H}_{i}^-=\left\{x\in{\rm I\!R}^{n_{xc}}\:\middle|\: l_i^Tx\leq\rho_{\mathcal{R}_{100}}(l_i)\right\}$---and 50 false---$\mathcal{H}_{i}^+=\left\{x\in{\rm I\!R}^{n_{xc}}\:\middle|\: l_i^Tx\geq1.1\rho_{\mathcal{R}_{100}}(l_i)\right\}$---results such that $\mathcal{R}_{100}\cap\mathcal{H}_i^-\not=\emptyset$ and $\mathcal{R}_{100}\cap\mathcal{H}_{i}^+=\emptyset$.

In Fig. \ref{fig-HalfandSupTimes}, the average computation times are similar between HCG and M2, which require solving MILPs of similar complexity. The computation time of M1 grows sharply in Case$(9,3)$, where the number of convex sets, and linear programs to be solved, jumps by two orders of magnitude. The reachable set given by the reduced hybrid zonotope, HCG-r, has the lowest computation times. However, this increased efficiency is at the cost of additional overhead in the generation of the set as shown in Table \ref{tab-HeatedRooms_HCGtimes}.

\section{Conclusions}\label{sec-conclusions}

Hybrid zonotopes extend zonotopes and constrained zonotopes to represent the nonconvex union of an exponential number of convex sets using a linear number of continuous and discrete variables. 
This is well-suited for reachability analysis of hybrid systems, in which discrete changes in dynamics can cause branching of sets. 
Furthermore, exact reachable sets of linear mixed logical dynamical systems can be calculated as a hybrid zonotope using set operations exhibiting linear growth in set representation complexity. Methods for the removal of redundant continuous factors, binary factors, and linear equality constraints of such reachable sets substantially reduced the set representation complexity in several numerical examples. 

\appendix
\bibliographystyle{unsrt} 
\bibliography{BibTex_2021}

\end{document}